\documentclass[letterpaper,twocolumn,prl,aps,superscriptaddress,floatfix]{revtex4}
\usepackage[latin1]{inputenc}
\usepackage{bm}
\usepackage{amssymb,amsbsy,amsmath}
\usepackage{graphicx}
\usepackage{verbatim}
\usepackage{color}
\usepackage{pifont}

\begin{document}

\title{Experimental observation of coherent-information superadditivity in a dephrasure channel}
\author{Shang Yu}
\affiliation{CAS Key Laboratory of Quantum Information, University of Science and
Technology of China, Hefei, 230026, China}
\affiliation{CAS Center For Excellence in Quantum Information and Quantum Physics,
University of Science and Technology of China, Hefei, 230026, P.R. China}
\author{Yu Meng}
\affiliation{CAS Key Laboratory of Quantum Information, University of Science and
Technology of China, Hefei, 230026, China}
\affiliation{CAS Center For Excellence in Quantum Information and Quantum Physics,
University of Science and Technology of China, Hefei, 230026, P.R. China}
\author{Raj B. Patel}
\email{raj.patel@physics.ox.ac.uk}
\affiliation{Clarendon Laboratory, Department of Physics, Oxford University, 
Parks Road OX1 3PU Oxford, United Kingdom}
\author{Yi-Tao Wang}
\affiliation{CAS Key Laboratory of Quantum Information, University of Science and
Technology of China, Hefei, 230026, China}
\affiliation{CAS Center For Excellence in Quantum Information and Quantum Physics,
University of Science and Technology of China, Hefei, 230026, P.R. China}
\author{Zhi-Jin Ke}
\affiliation{CAS Key Laboratory of Quantum Information, University of Science and
Technology of China, Hefei, 230026, China}
\affiliation{CAS Center For Excellence in Quantum Information and Quantum Physics,
University of Science and Technology of China, Hefei, 230026, P.R. China}
\author{Wei Liu}
\affiliation{CAS Key Laboratory of Quantum Information, University of Science and
Technology of China, Hefei, 230026, China}
\affiliation{CAS Center For Excellence in Quantum Information and Quantum Physics,
University of Science and Technology of China, Hefei, 230026, P.R. China}
\author{Zhi-Peng Li}
\affiliation{CAS Key Laboratory of Quantum Information, University of Science and
Technology of China, Hefei, 230026, China}
\affiliation{CAS Center For Excellence in Quantum Information and Quantum Physics,
University of Science and Technology of China, Hefei, 230026, P.R. China}
\author{Yuan-Ze Yang}
\affiliation{CAS Key Laboratory of Quantum Information, University of Science and
Technology of China, Hefei, 230026, China}
\affiliation{CAS Center For Excellence in Quantum Information and Quantum Physics,
University of Science and Technology of China, Hefei, 230026, P.R. China}
\author{Wen-Hao Zhang}
\affiliation{CAS Key Laboratory of Quantum Information, University of Science and
Technology of China, Hefei, 230026, China}
\affiliation{CAS Center For Excellence in Quantum Information and Quantum Physics,
University of Science and Technology of China, Hefei, 230026, P.R. China}
\author{Jian-Shun Tang}
\email{tjs@ustc.edu.cn}
\affiliation{CAS Key Laboratory of Quantum Information, University of Science and
Technology of China, Hefei, 230026, China}
\affiliation{CAS Center For Excellence in Quantum Information and Quantum Physics,
University of Science and Technology of China, Hefei, 230026, P.R. China}
\author{Chuan-Feng Li}
\email{cfli@ustc.edu.cn}
\affiliation{CAS Key Laboratory of Quantum Information, University of Science and
Technology of China, Hefei, 230026, China}
\affiliation{CAS Center For Excellence in Quantum Information and Quantum Physics,
University of Science and Technology of China, Hefei, 230026, P.R. China}
\author{Guang-Can Guo}
\affiliation{CAS Key Laboratory of Quantum Information, University of Science and
Technology of China, Hefei, 230026, China}
\affiliation{CAS Center For Excellence in Quantum Information and Quantum Physics,
University of Science and Technology of China, Hefei, 230026, P.R. China}

\begin{abstract}
We present an experimental approach to construct a \emph{dephrasure} channel, which contains both dephasing and erasure noises, and can be used as an efficient tool to study the superadditivity of coherent information. By using a three-fold dephrasure channel, the superadditivity of coherent information is observed, and a substantial gap is found between the zero single-letter coherent information and zero quantum capacity. Particularly, we find that when the coherent information of $n$ channel uses is zero, in the case of larger number of channel uses, it will become positive. These phenomena exhibit a more obvious superadditivity of coherent information than previous works, and demonstrate a higher threshold for non-zero quantum capacity. Such novel channels built in our experiment also can provide a useful platform to study the non-additive properties of coherent information and quantum channel capacity.
\end{abstract}

\maketitle
\date{\today}
\textit{Introduction.}---Channel capacity is at the heart of information theory and is used to quantify the ability of a practical communication channel to transmit the information from a sender to a receiver. Regarding the quantum channel $\varepsilon$, the notion of channel capacity has multiple formulations in different settings: classical capacity $C(\varepsilon)$~\cite{Schumacher1997,Holevo1998}, which denotes the ability of the channel to deliver classical information; entanglement-assisted classical capacity $C_{E}(\varepsilon)$~\cite{Bennett1996,Bennett1999}, which shows the same ability when the sender and receiver share some entanglement resources; and private capacity $P(\varepsilon)$~\cite{Devetak2005}, which describes the maximum classical bit that can be transmitted while negligible information can be intercepted by third parties. In quantum communication theory, the ability to convey quantum information is denoted as the quantum capacity $Q(\varepsilon)$~\cite{Seth1997,DiVincenzo1998,Macchiavello2016,Leditzky2018}, which is defined by the maximum of coherent information $I_{c}$. Many recent studies focused on exploring various properties of quantum capacity~\cite{Gyongyosi2018}: for example, the lower bounds of quantum capacity in a noisy quantum channel have been detected~\cite{Macchiavello2016,Cuevas2017}, and the upper bounds of quantum capacity are derived in thermal attenuator channels~\cite{Rosati2018}.

The amount of coherent information present in a quantum channel plays a central role in quantum communication. The quantum capacity of a channel~\cite{Gyongyosi2018} can be computed using the methods of Lloyd, Shor and Devetak~\cite{Seth1997,Shor2002,Devetak2003}, while coherent information itself can be considered analogous to the role of mutual information in classical channels. In contrast to the mutual information, which is additive in classical channels, coherent information exhibits superadditivity. Consequently, the quantum capacity cannot be computed by a single optimization like the classical one~\cite{Cubitt2015}, which makes the measurements of quantum capacity challenging. For this reason, the superadditive properties of coherent information becomes important in computing the quantum capacity~\cite{Smith2008,Hastings2009}. The first example of superadditivity is given by DiVincenzo \emph{et al.}~\cite{DiVincenzo1998}, and then has been extended by Smith \emph{et al.}~\cite{Smith2007}, in which they provide a heuristic for designing codes for general channels. Just recently, Cubitt \emph{et al.}~\cite{Cubitt2015} find that for any number of channel uses, channels exist for which there is positive capacity and zero coherent information.

These additive-violated results indicate our limited understanding of the superadditivity of coherent information, and the limitations in studying the quantum capacity. These limitations come down to the fact that we do not have a useful tool to exhibit the superadditive phenomenon of coherent information, especially in an experimental testing. Therefore, to understand the coherent information and, directly, the quantum capacity more efficiently and profoundly, a family of quantum channels that can exhibit non-additivity is urgently needed.

In this letter, we build a \emph{dephrasure} channel~\cite{Leditzky2018} that can exert both the dephasing and the erasure evolution on the input state. Assisted by this antidegradable channel, the superadditivity of coherent information has been studied. By coding the information with different numbers of photons and applying the corresponding number of channel uses, we find that a substantial gap of coherent information exists between the $n$- and $(n+1)$-letter level. This evidence well demonstrates the superadditive property of the coherent information and also shows a channel with zero coherent information can have positive capacity~\cite{Cubitt2015}. The results obtained in our study provide a deeper understanding of the nonadditivity of coherent information, and supply a solid basis for inspiring many new ideas on coherent information and quantum capacity.

\begin{figure*}[tb]
\centering
\includegraphics[width=0.76\textwidth]{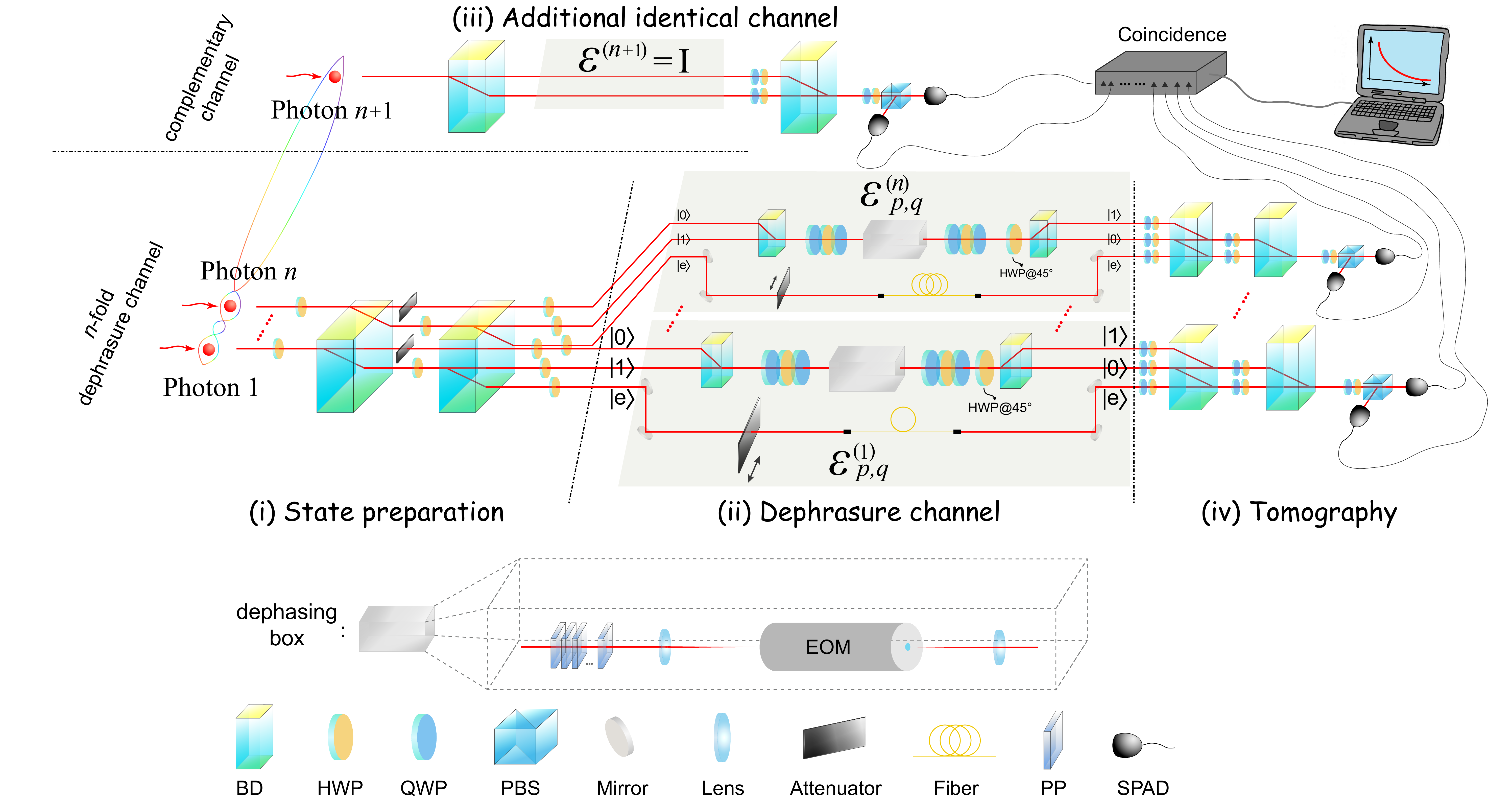}
\caption{Experimental setup consisting of four parts: (i) state preparation [the $\beta$-barium borate (BBO) crystals and the pump laser are not presented in the figure]; (ii) a three-fold dephrasure channel, which will induce a dephasing evolution between $|0\rangle$ and $|1\rangle$ bases in the gray dephasing box via the PPs-EOM sets, and cause an erasure evolution by adding a flag basis $|e\rangle$ via adjustment of an attenuator; (iii) additional identical channel platform, which receives the ($n+1$)th photon and can prepare the repetition code remotely and, meanwhile, also acts as an additional path for building the complementary channel $\varepsilon^{c}$; (iv) detection devices. Keywords include: HWP, half-wave plate; QWP, quarter-wave plate; PBS, polarizing beam splitter; PP, phase plate; EOM, electro-optic modulator; BD, beam displacer; SPAD, single-photon avalanche diode.}
\label{Fig1}
\end{figure*}

\textit{Coherent information, dephrasure channel, and experimental setup.}--- The quantum capacity $Q(\varepsilon)$, which describes the capability of the channel to convey quantum information, can be quantified by the maximum of coherent information. The quantity of coherent information $I_{c}$, depends on the properties of the channel, and is also strongly related to the form of error-correcting code used in the communication~\cite{DiVincenzo1998,Hastings2009}. Thus, the additive property of $I_{c}$ will also vary accordingly~\cite{Leditzky2018}. The quantum capacity of a noisy channel $\varepsilon$ can be expressed as~\cite{Seth1997,Barnum1998,Devetak2005}
\begin{equation}
\begin{aligned}
Q(\varepsilon):=\lim_{n\to\infty}Q_{n}(\varepsilon),
\end{aligned}
\end{equation}
where $Q_{n}(\varepsilon)=\frac{1}{n}\max_{\rho}\{I_{c}(\rho,\varepsilon_{n})\}$, and $I_{c}(\rho,\varepsilon_{n})=S(\varepsilon_{n}(\rho))-S[\varepsilon_{n}^{c}(\rho))]$ expresses the coherent information of the channel~\cite{Schumacher1996,Barnum2000}. Here, $S(\cdot)$ is the von Neumann entropy, and $\varepsilon^{c}$ denotes the complementary channel of $\varepsilon$. Clearly, to study the quantum capacity, the properties of coherent information are an important target to conduct research and exploration, especially superadditivity. However, studies of the superadditivity of coherent information are very limited~\cite{Leditzky2018}, which causes difficulties in understanding quantum capacity.

To address this challenge, we construct a dephrasure channel with a different number of uses (Fig. 1, middle part). This channel contains both dephasing~\cite{Nielsen2000} and erasure~\cite{Grassl1997,Bennett1997} noises, and can be used to study the superadditivity of coherent information since its antidegradability. According to Ref.~\cite{Leditzky2018}, this dephrasure channel can be expressed as:
\begin{equation}
\begin{aligned}
\varepsilon_{p,q}(\rho)=(1-q)[(1-p)\rho+p\Pi\rho\Pi^{\dagger}]+q\text{Tr}(\rho)|e\rangle\langle e|,
\end{aligned}
\end{equation}
where $p$ and $q$ ($p,q\in[0,0.5]$) denote the dephasing and erasure noise strength in the channel, $\Pi=\bf{n}\cdot\hat{\sigma}$ ($\hat{\sigma}$ are the Pauli operators), and $|e\rangle$ denotes the erasure flag basis which is orthogonal to the input space. This channel can be efficiently realized with an optical setup by controllable qutrit evolution, and the error-correcting codes can be prepared by a multiphoton entangled source.

Fig. 1 presents our experimental setup, which can be recognized as four parts: (i) the state preparation platform, which contains a multiphoton-entangled source~\cite{zhang2016} (not presented in the figure); (ii) three-fold dephrasure channel; (iii) an additional identical channel, which acts as a complementary channel as well as remote state preparation; and (iv) tomography device for the qutrit state. In our experiment, we use spontaneous parametric down conversion process to create a four-photon GHZ state as $|\Phi^{+}\rangle$ = $(1/\sqrt{2})(|HHHH\rangle+|VVVV\rangle)$~\cite{zhang2016}, which can be used to construct the repetition codes applied in the communication, and the corresponding purification codes for coherent information measurements (the details of the photon source can be found in Supplemental Material~\cite{sm}). The dephrasure channel [$\varepsilon_{p,q}(\cdot)$] is constituted by two types of sub-channels: dephasing and erasure channels. In the dephasing channel, the input qubit state, $1/\sqrt{2}(|0\rangle+|1\rangle)$ ($|0\rangle$ and $|1\rangle$ are the two orthogonal bases in the path degree of freedom), will suffer a dephasing evolution. It is realized by a pair of BD, two sets of waveplates~\cite{Yu2017,wpg}, and a dephasing box that contains a PPs-EOM set. This structure converts the path into the polarization degree of freedom and can achieve the dephasing to a very small degree as well as at arbitrary directions. By adding the PP into the channel and adjusting the voltage on the EOM(see  Supplemental Material~\cite{sm} for the details), we can control the degree of decoherence (represented by $p$) of the state. The erasure channel will add a flag state $|e\rangle$ into the input state, which transforms the original qubit state into a qutrit state. It is realized by adjusting the proportion of photons residing in the flag basis $|e\rangle$ through shifting the variable attenuator in the experiment. This allows us to change the erasure strength $q$ from $0$ to $0.5$.  A 0.5-m fiber in the path of $|e\rangle$ ensures that the flag state merges into the input qubit state incoherently, which satisfies the evolution expressed by Eq.~(2). In the experiment, we build a three-fold dephrasure channel and let each photon go through the channel simultaneously. An additional photon and the identity channel $\varepsilon^{(n+1)}=I$ are used to build the complementary channel $\varepsilon_{n}^{c}=\varepsilon_{p,q}^{(1)}\otimes\cdots\otimes\varepsilon_{p,q}^{(n)}\otimes I$, which will be used in the coherent information measurements. The final state is obtained by qutrit state-tomography process using QWP and HWP pairs, 2 BDs, PBS, and SPADs. Then, the coherent information of the channel can be extracted from the state-tomography results of the final state $\varepsilon(\rho)$ and $\varepsilon^{c}(\rho)$.

\begin{figure}[tb]
\includegraphics[width=0.47\textwidth]{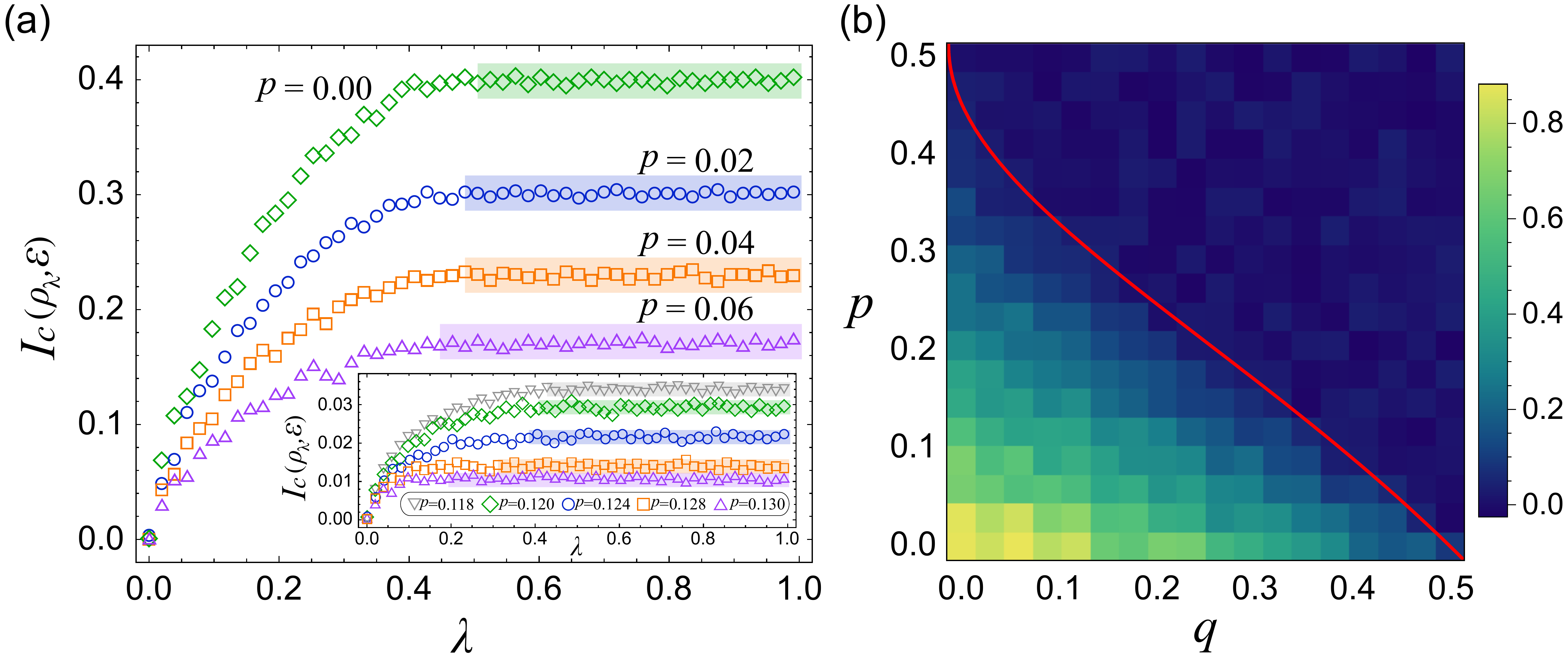}
\caption{Experimental results of the single-letter coherent information $I_{c}(\varepsilon)$. (a) Single-letter coherent information measurements. By searching ${\lambda}$, the optimal $I_{c}(\rho_{\lambda},\varepsilon)$ can be determined because $I_{c}(\varepsilon)$ $=$ $\max_{\rho_{\lambda}}I_{c}(\rho_{\lambda},\varepsilon)$. The colored areas indicate the convergent region. (b) Density plot of $I_{c}(\varepsilon)$ with various $p$, and $q$. Optimal $\lambda$ was determined in the experiment for each value of $I_{c}(\varepsilon)$. Red contour denotes the boundary of $I_{c}(\varepsilon)>0$.}
\label{Fig2}
\end{figure}

\textit{Experimental results for coherent information in dephrasure channel.}---Figure 2 presents the experimental results of single-letter coherent information $I_{c}(\varepsilon)$. For simplicity and without loss of generality, in our experiment, the dephrasure channel is fixed along the $Z$ dephasing direction, i.e., $\Pi=\hat{\sigma}_{z}$. Thus, the best choice of $\rho$ has the form of $\lambda|0\rangle\langle0|+(1-\lambda)|1\rangle\langle1|$ where $\lambda\in[0,1]$, because this state will not suffer from the $Z$ dephasing~\cite{Leditzky2018}. To measure the $I_{c}(\varepsilon)$, two entangled photons have to be used in the experiments. First, one of the photons, which is prepared as $\rho$, is sent into the dephrasure channel (denoted by $\varepsilon_{p,q}^{(1)}$ in Fig. 1). Since the two photons are entangled, the abovementioned photon state can be remotely prepared by the second photon, which will be sent into the complementary channel (see Supplemental Materials for details~\cite{sm}). Then, its entropy can be read out by state tomography at the end. As for the complementary term $S[\varepsilon_{p,q}^{c}(\rho)]$, we can obtain it from a purified state $\rho'$~\cite{DiVincenzo1998,Leditzky2018}, where $\rho'=|\phi\rangle\langle\phi|$, and $|\phi\rangle=\sqrt{\lambda}|0\rangle|0\rangle+\sqrt{1-\lambda}|1\rangle|1\rangle$. Therefore, the measurements of $S[\varepsilon_{p,q}^{c}(\rho)]$ map to measurements of $S[\varepsilon^{(1)}_{p,q}\otimes\mathbb{I}(\rho')]$ ($\varepsilon^{(2)}=\mathbb{I}$ is an identity channel, presented at the top of Fig. 1). Similarly, the entropy can be obtained by the bipartite tomography results.

\begin{figure}[tb]
\includegraphics[width=0.47\textwidth]{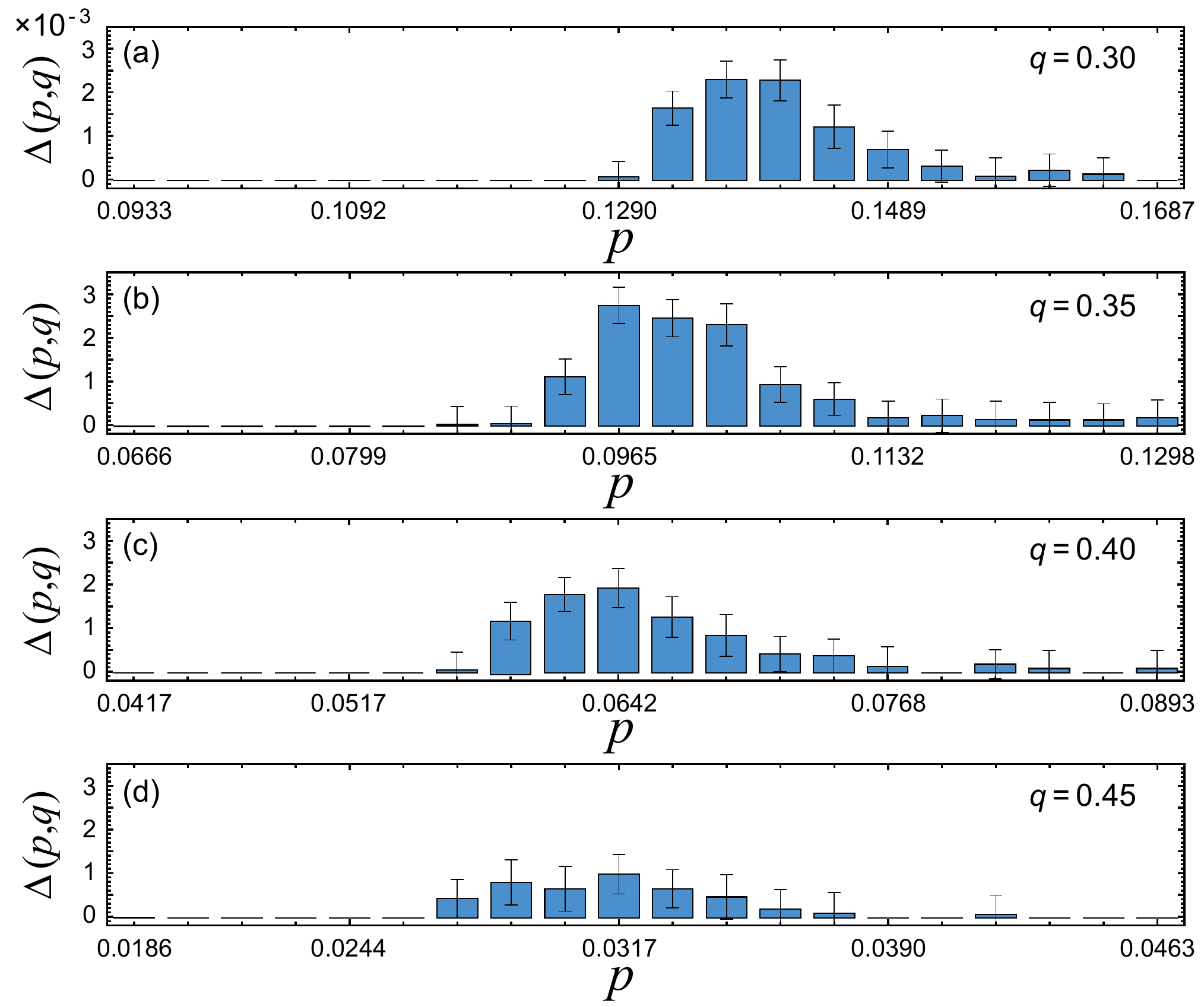}
\caption{Experimental results of coherent information superadditivity. Under various channel parameters, we measure the value $\Delta(p,q)$ which expresses the distance of $Q_{n}(\varepsilon)$ between the two-letter and single-letter levels; the non-positive values are set to zero, as shown in (a)$\sim$(d). The non-zero blue bars presented in this figure demonstrate well the non-additivity of coherent information. The error bars are calculated through Monte Carlo simulations which consider the counting noise (the same below without declaration).}
\label{Fig3}
\end{figure}

To obtain $I_{c}(\varepsilon)$, we have to determine an optimal input state (i.e., the optimal value of $\lambda$) for communication under the specific channel parameters. In Fig. 2(a), we measure the coherent information with various $\lambda$ and $p$ when the erasure strength is set as $q=0.3$. It is easy to check that when $\lambda$ increases, the coherent information converges at approximately $\lambda=0.5$. While the dephasing strength becomes larger, the coherent information converges at a relative early stage; see the inset of Fig. 2(a). Fig. 2(b) shows the coherent information obtained under various channel parameters $p$ and $q$. It is clear that coherent information decays along with the increasing dephasing and erasure strength. The red curve [$q<\frac{(1-2p)^{2}}{1+(1-2p)^{2}}$] denotes the boundary for $I_{c}(\varepsilon)\ge0$~\cite{Leditzky2018}. Although these results are not the main objective in this study, the data obtained under the single-letter level paves the way for the following studies.

\textit{Testing the non-additivity of coherent information.}---The most interesting aspect of the dephrasure channel is the superadditivity of coherent information, which means that a higher communication rate can be achieved when $n\geq2$ than previously expected. To obtain a higher coherent information, the repetition code has to be used in the communication. In this experiment, we use the weighted repetition code $\rho_{n}=\lambda|0\rangle\langle0|^{\otimes n}+(1-\lambda)|1\rangle\langle1|^{\otimes n}$ ($n\geq2$) for communication and demonstrate that under multi-letter level the coherent information exhibits the superadditive property. First, we put our focus on the two-letter level ($n=2$). In this situation, three entangled photons have to be used: two-photon maximally mixed state is used to measure $S(\rho_{2},\varepsilon_{2})$; three entangled photons are used as the state $\rho'=|\phi\rangle\langle\phi|$ (where $|\phi\rangle=\sqrt{\lambda}|000\rangle+\sqrt{1-\lambda}|111\rangle$) to measure $S(\rho'_{2},\varepsilon_{2}^{c})$ ($\varepsilon_{2}^{c}=\varepsilon_{p,q}^{(1)}\otimes\varepsilon_{p,q}^{(2)}\otimes\mathbb{I}$). Here, we use the quantity $\Delta$ to verify the non-additivity of coherent information between the single- and two-letter levels, i.e., $\Delta(p,q)$ = $Q_{2}(\varepsilon)-Q_{1}(\varepsilon)=$$\max_{\rho}\{\frac{1}{2}I_{c}(\rho_{2},\varepsilon_{2})$ $-$ $I_{c}(\varepsilon)\}$. Since the maximal $\Delta(p,q)$ is contained in the region $\frac{1-2p-2p(1-p)\ln\frac{1-p}{p}}{2-2p-2p(1-p)\ln\frac{1-p}{p}}<q<\frac{(1-2p)^{2}}{1+(1-2p)^{2}}$~\cite{Leditzky2018}, for simplicity, we can perform the measurements in this area. In Fig. 3, we plot the positive part of $\Delta$ for various $p$, and $q$. The physical meaning of $\Delta(p,q)$ is clear: it provides a measure of how greater the communication rate is in the two-letter case compared to the single-letter case. This phenomenon clearly demonstrates the superadditivity of coherent information, i.e., $Q_{n}(\varepsilon_{n})\ge I_{c}(\varepsilon)$. Figures 3 (a)-(d) shows that when the erasure strength is light (a relatively small $q$), the non-additivity phenomenon happens at a relatively large dephasing strength (a relatively large $p$). The non-additive peak will move to the left along with an increasing $q$. During this process, we can find a pair of specific channel parameters to obtain the largest non-additivity value $\Delta(p,q)$. From Figs. 3 (a) and (b), we determined that the optimal $p$, $q$ are located at $p\in[0.09,0.15]$, and $q\in[0.30,0.35]$. To accurately determine the optimal parameters, we experimentally measure $\Delta(p,q)$ in a more detailed area (see Supplemental Material for the results~\cite{sm}), and find that the largest $\Delta(p,q)$ is located at $q\simeq3p$~\cite{Leditzky2018}.

We now study the superadditivity of coherent information in multi-letter situations, where we extend our exploration up to the three-letter level using four entangled photons. On the basis of the abovementioned results, for simplicity, we fix the erasure strength to $q=3p$, and vary the dephasing strength $p$ from 0.1115 to 0.1195, as shown in Fig. 4. In this region, we determine that the coherent information measured under $n=2$ (denoted by blue circles), and $n=3$ (denoted by orange squares) always exceeds the quantities under single-letter level (denoted by gray triangles). Therefore, we can conclude that in the dephrasure channel, the quantum capacities under multi-level cases indeed exhibit their advantages. In addition, in the inset in Fig. 4, we show an extreme situation in which the coherent information under multi-letter levels will exist even when zero coherent information exists in the single-letter case. To reveal this phenomenon, we measured in detail the region $p\in[0.1185,0.1200]$. By adjusting the voltage added to the EOM in the dephasing box, we can fine-tune the dephasing strength $p$, and obtain the results which show that coherent information is still present under the multi-letter cases, where none exists in the single-letter case. What's more, we also find that while coherent information ceases to exist in the two-letter case, it remains present in the three-letter case. This evidence proves that for the channel with unbounded number of uses, the coherent information of $n$-letter level is zero, but in the $(n+1)$-letter level case, the coherent information is strictly positive. In the previous examples, two channel uses are already sufficient. This phenomenon observed in experiments provides a strong evidence to refute the previously understood, and demonstrates that a channel with zero coherent information can have a positive capacity~\cite{Cubitt2015}.


\begin{figure}[tb]
\includegraphics[width=0.47\textwidth]{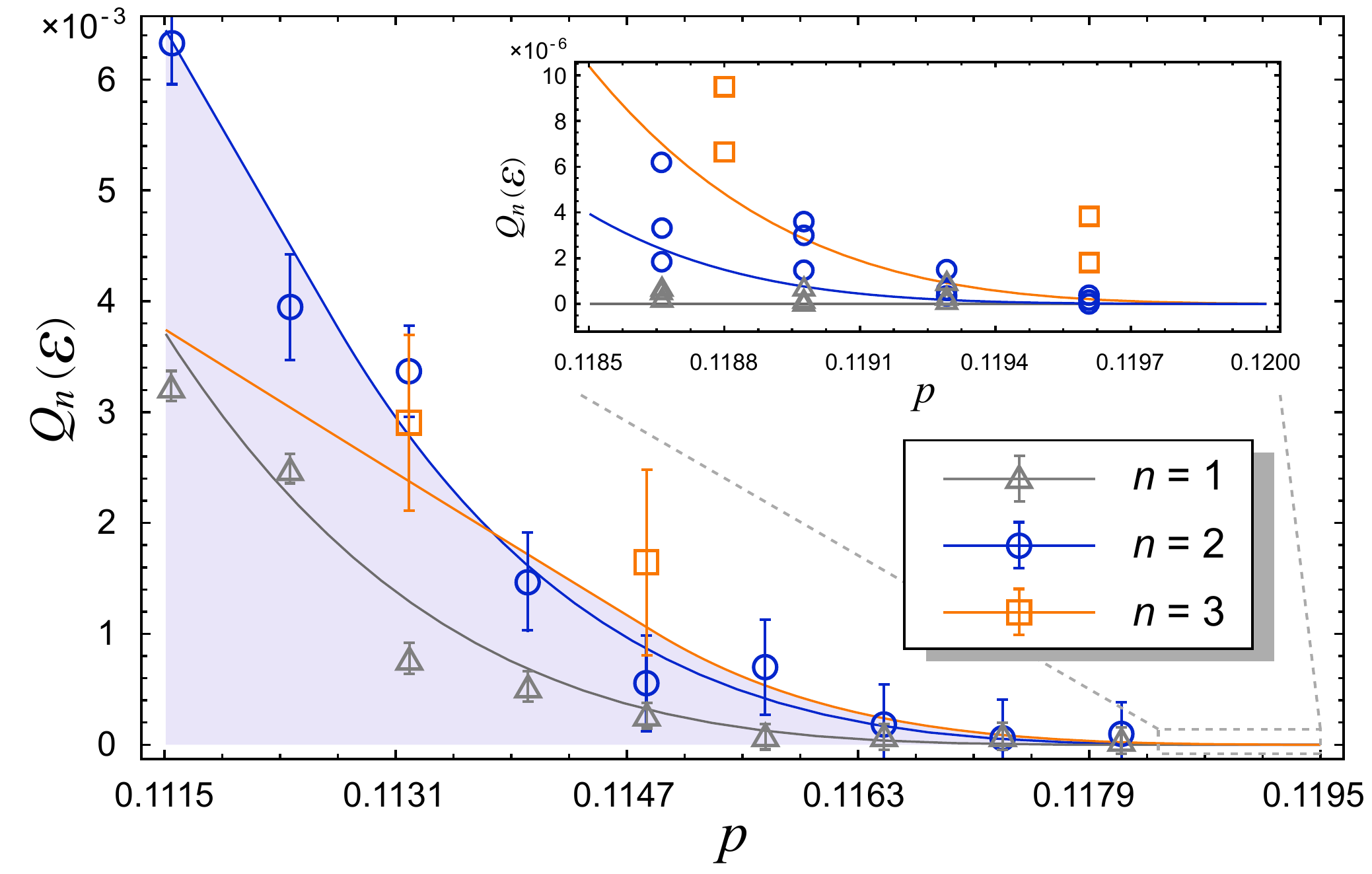}
\caption{$Q_{n}(\varepsilon)$ with $n$-photon repetition code and $n$ channel uses. In our experiment, we extend this exploration to $\rho_{3}$ and $\varepsilon_{3}$ with four entangled photons. The gray triangles, blue circles, and orange squares indicate the experimental results, and the lines indicate their corresponding theoretical value. For the use of the repetition code (both $n=2$, and $n=3$), the advantages are clearly shown compared with the single-letter level. The shaded area shows the supremum of coherent information under $n$ ($n=1,2,3$) channel uses. The inset shows the superadditivity of $Q_{n}(\varepsilon)/n$ when the single-letter coherent information is zero.}
\label{Fig4}
\end{figure}

\textit{Discussion and conclusion.}---Besides the quantum codes used in our experiments, it is interesting to study some codes that have entanglement, because the nonlocal correlations between the photons can resist part of the noise, and may have advantage in some certain situations~\cite{Giovannetti2006,Hastings2009,Demkowicz2014,Huang2016}. Recently, neural network coding form~\cite{Bausch2020} also shows the potential ability to yield high coherent information. Furthermore, we can see that a larger number of channel uses is necessary for detecting positive coherent information. An interesting phenomenon is to detect a non-zero coherent information at $(n+1)$-letter level, while $n$$\sim$single-letter level is zero. Namely, $Q_{n}(\varepsilon)=0$ but $Q(\varepsilon)>0$~\cite{Cubitt2015}. In such a situation, multi-qubit entangled states are needed in the tests, and this requires additional SPDC sources~\cite{Wang2016}, or other encoding forms, like time-bin encoding~\cite{Schon2005,Takesue2009}.

In our experiments, we applied various repetition codes and different numbers of channel uses to detect the coherent information in a dephrasure channel, which has a ``dephasing $+$ erasure'' structure and is useful to study the non-additivity. Based on the observed results, we find a significant difference between the zero $n$-letter and zero $(n+1)$-letter coherent information (up to three-letter level, i.e., three channel uses). Namely, the coherent information for $n$ uses fails to exhibit the dephrasure channel has quantum capacity, but the  channel capacity can be disclosed by measuring the $I_{c}$ under a larger number of channel uses. These evidences confirm the superadditivity of coherent information, and a direct corollary is that a finite number of channel uses is not sufficient for measuring the quantum channel capacity. This work provides the first demonstration of a dephrasure channel. Our reliable data pave the way for the future study of the non-additivity of coherent information, quantum capacity, as well as this newfashioned \emph{dephrasure} channel.

\textbf{Acknowledgments}
This work is supported by the National Key Research and Development Program of China (No. 2017YFA0304100), the National
Natural Science Foundation of China (Grants Nos. 61327901, 11674304, 11822408, 11774335, and 11821404), the Key Research Program of Frontier Sciences of the Chinese Academy of Sciences (Grant No. QYZDY-SSW-SLH003), the Youth Innovation Promotion Association of Chinese Academy of Sciences (Grants No. 2017492), the Foundation for Scientific Instrument and Equipment Development of Chinese Academy of Sciences (No. YJKYYQ20170032), the research was funded by Fok Ying Tung Education Foundation (No. 171007),Science Foundation of the CAS (No. ZDRW-XH-2019-1), Anhui Initiative in Quantum Information Technologies (AHY020100, AHY060300), the National Postdoctoral Program for Innovative Talents (Grant No. BX20180293), China Postdoctoral Science Foundation funded project (Grant No. 2018M640587), the Fundamental Research Funds for the Central Universities (No. WK2470000026).

S. Yu and Y. Meng contributed equally to this work.


\begin{thebibliography}{38}
\bibitem{wpg}
The wave-plate group is composite by quarter, half and quarter-wave plates. Such a three-wave-plate group can realize the arbituary single qubit unitary evolotion.
\end{thebibliography}
\end{document}


\title{Supplemental Material: Experimental observation of coherent-information superadditivity in a dephrasure channel}
\author{Shang Yu}
\affiliation{CAS Key Laboratory of Quantum Information, University of Science and
Technology of China, Hefei, 230026, China}
\affiliation{CAS Center For Excellence in Quantum Information and Quantum Physics,
University of Science and Technology of China, Hefei, 230026, P.R. China}
\author{Yu Meng}
\affiliation{CAS Key Laboratory of Quantum Information, University of Science and
Technology of China, Hefei, 230026, China}
\affiliation{CAS Center For Excellence in Quantum Information and Quantum Physics,
University of Science and Technology of China, Hefei, 230026, P.R. China}
\author{Raj B. Patel}
\email{raj.patel@physics.ox.ac.uk}
\affiliation{Clarendon Laboratory, Department of Physics, Oxford University, 
Parks Road OX1 3PU Oxford, United Kingdom}
\author{Yi-Tao Wang}
\affiliation{CAS Key Laboratory of Quantum Information, University of Science and
Technology of China, Hefei, 230026, China}
\affiliation{CAS Center For Excellence in Quantum Information and Quantum Physics,
University of Science and Technology of China, Hefei, 230026, P.R. China}
\author{Zhi-Jin Ke}
\affiliation{CAS Key Laboratory of Quantum Information, University of Science and
Technology of China, Hefei, 230026, China}
\affiliation{CAS Center For Excellence in Quantum Information and Quantum Physics,
University of Science and Technology of China, Hefei, 230026, P.R. China}
\author{Wei Liu}
\affiliation{CAS Key Laboratory of Quantum Information, University of Science and
Technology of China, Hefei, 230026, China}
\affiliation{CAS Center For Excellence in Quantum Information and Quantum Physics,
University of Science and Technology of China, Hefei, 230026, P.R. China}
\author{Zhi-Peng Li}
\affiliation{CAS Key Laboratory of Quantum Information, University of Science and
Technology of China, Hefei, 230026, China}
\affiliation{CAS Center For Excellence in Quantum Information and Quantum Physics,
University of Science and Technology of China, Hefei, 230026, P.R. China}
\author{Yuan-Ze Yang}
\affiliation{CAS Key Laboratory of Quantum Information, University of Science and
Technology of China, Hefei, 230026, China}
\affiliation{CAS Center For Excellence in Quantum Information and Quantum Physics,
University of Science and Technology of China, Hefei, 230026, P.R. China}
\author{Wen-Hao Zhang}
\affiliation{CAS Key Laboratory of Quantum Information, University of Science and
Technology of China, Hefei, 230026, China}
\affiliation{CAS Center For Excellence in Quantum Information and Quantum Physics,
University of Science and Technology of China, Hefei, 230026, P.R. China}
\author{Jian-Shun Tang}
\email{tjs@ustc.edu.cn}
\affiliation{CAS Key Laboratory of Quantum Information, University of Science and
Technology of China, Hefei, 230026, China}
\affiliation{CAS Center For Excellence in Quantum Information and Quantum Physics,
University of Science and Technology of China, Hefei, 230026, P.R. China}
\author{Chuan-Feng Li}
\email{cfli@ustc.edu.cn}
\affiliation{CAS Key Laboratory of Quantum Information, University of Science and
Technology of China, Hefei, 230026, China}
\affiliation{CAS Center For Excellence in Quantum Information and Quantum Physics,
University of Science and Technology of China, Hefei, 230026, P.R. China}
\author{Guang-Can Guo}
\affiliation{CAS Key Laboratory of Quantum Information, University of Science and
Technology of China, Hefei, 230026, China}
\affiliation{CAS Center For Excellence in Quantum Information and Quantum Physics,
University of Science and Technology of China, Hefei, 230026, P.R. China}
\maketitle
\date{\today}

\section{Details of the photon source, and the PPs-EOM set}
The photons for encoding the input state come from the spontaneous parametric down conversion (SPDC) process, same as the setup shown in Ref.~\cite{zhang2016}. The photons then pass through 3 nm bandpass filters. In order to determine the degree of dephasing under different relative delay, we measure the dynamical property of the decoherence. The results are shown in Fig. S1. We find when we add 392$\lambda$ relative delay, the degree of coherence is already reduced to 0.0011. Then we fit these data with a Gaussian-decay model, $a\times e^{-\frac{\lambda^{2}}{\omega^{2}}}$. According to the data at hand, the fitted values are: $a=0.986$, and $\omega=1.357\times10^{-4}$. Then, we can map the noise parameter $p$ to the corresponding dephasing degree, and then, a specific relative delay.
\begin{figure}[tb]
\includegraphics[width=0.47\textwidth]{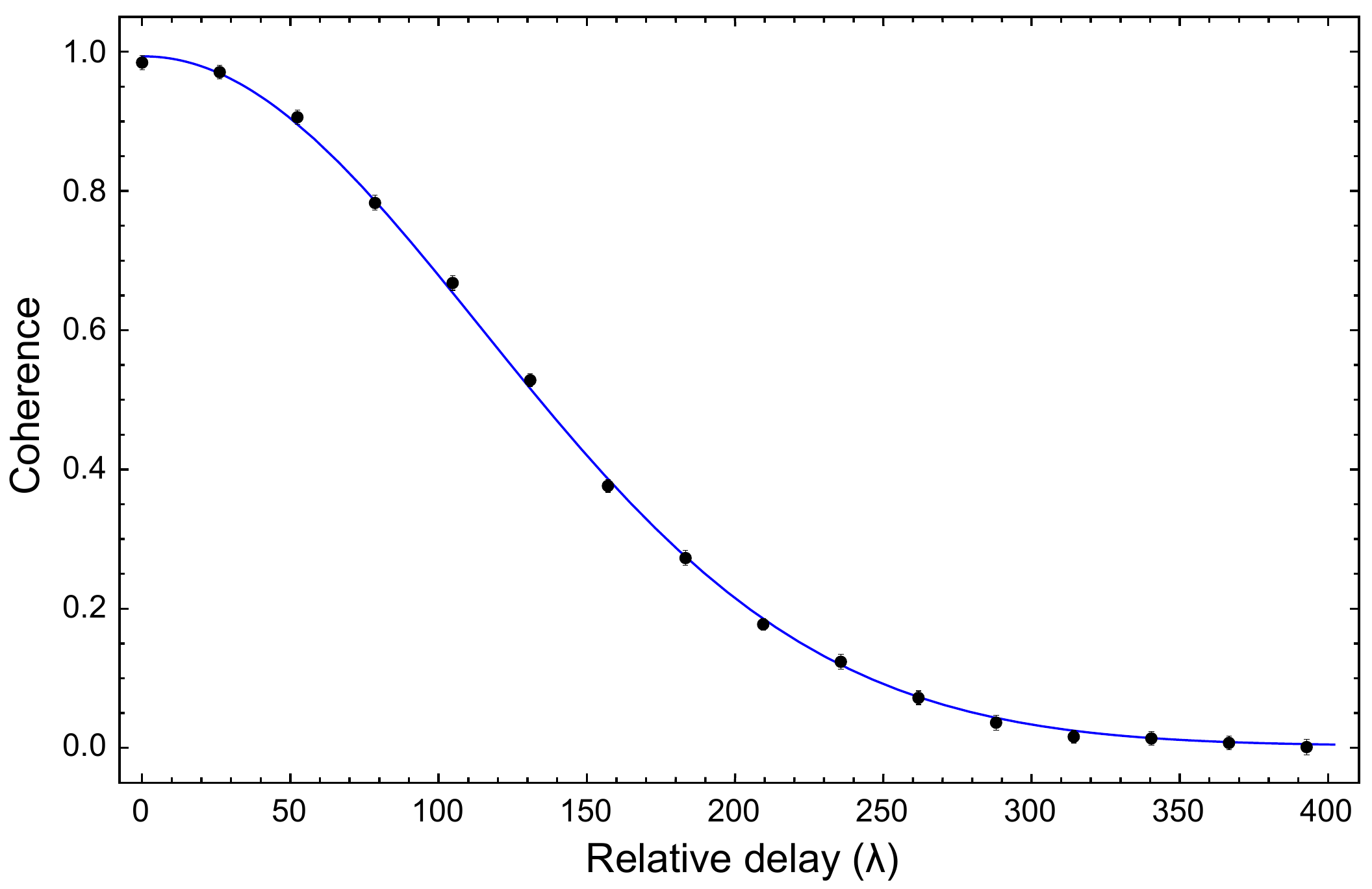}
\renewcommand{\thefigure}{S1}
\caption{\label{FigS1}
Results of the dynamical property of decoherence. The black dots are the experimental results, and the blue line is the theoretical fitting result.}
\label{FigS1}
\end{figure}

For adjusting the noise parameter $p$ accurately, we use a PPs-EOM set. The phase plates can coarse control the dephasing degree. However, in this experiment, we need to perform a very small change in $p$. For example, in the inset of Fig. 4, the smallest change is $\Delta p=0.0003$. This involves a small phase change within $2\pi$. Although some ordinary methods, like tilting a pair of phase plate, can also achieve this change, the accuracy and stability are not enough. Because of this, we apply an EOM (ConOptics Model 360) to achieve this by varying the applied voltage. EOM can realize a fine adjustment on the relative delay within $2\pi$. The test results are shown in Fig. S2. From the results, we find the first maximal photon count is located at -30 V. Through further increase in the voltage (going through 2$\pi$ phase change), the next maximal photon count locates at 125 V. Thus, we can see that changing 155 V on EOM can add 2$\pi$ (one $\lambda$) relative delay. The adjustment accuracy of EOM is 1 V, so the minimum adjustment accuracy will be $1/155\lambda$, namely, as for parameter $p$, the minimum adjustment accuracy is $\Delta p_{\text{min}}=$8.229$\times 10^{-6}$. Therefore, the PPs-EOM set can achieve the requirement.

\begin{figure}[tb]
\includegraphics[width=0.47\textwidth]{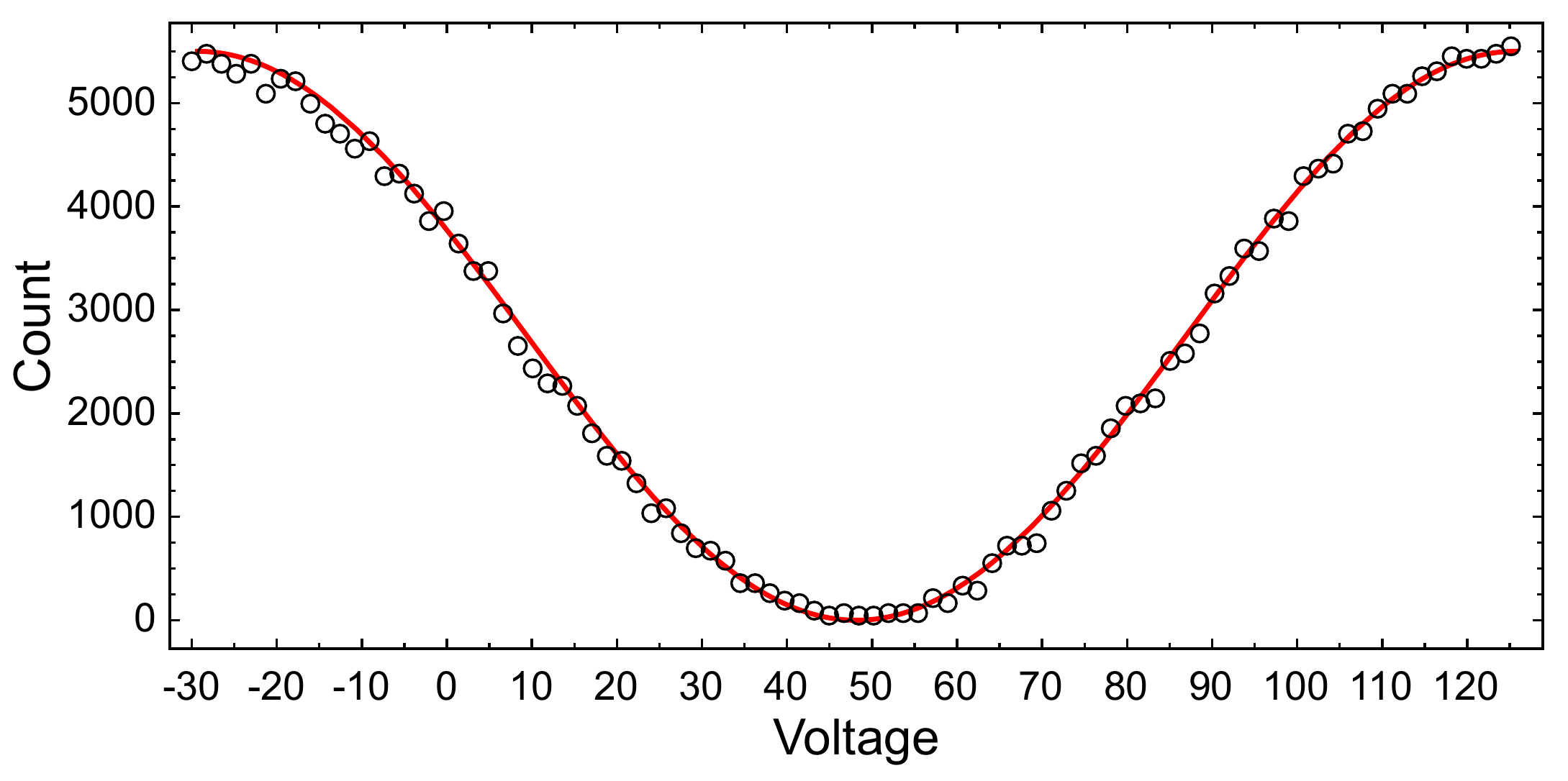}
\renewcommand{\thefigure}{S2}
\caption{\label{FigS2}
Voltage added on EOM versus photon counts. The black hollow dots are the experimental results, and the red line is the theoretical value $\frac{1}{2}$($1+\cos[{v-v_{0}}/v_{\frac{\lambda}{2}}\times\pi]$), where $v_{0}$ is the bias voltage, and $v_{\frac{\lambda}{2}}$ is the half wave voltage voltage.}
\label{FigS2}
\end{figure}

\section{Remotely preparing the input state}
The input photon state is initially prepared as the GHZ state $|\Phi^+\rangle=\frac{1}{\sqrt{2}}(|H\rangle^{\otimes n}+|V\rangle^{\otimes n})$, which can be rewritten as $|\Phi^+\rangle=\frac{1}{\sqrt{2}}(|\psi\rangle|\psi'\rangle+|\psi^\perp\rangle|\psi'^\perp\rangle)$ with $|\psi\rangle$ representing an arbitrary pure $n-1$ qubit state and $|\psi'\rangle$ representing the state of the last photon. $|\psi^\perp\rangle$ and $|\psi'^\perp\rangle$ represents the corresponding orthogonal states. When the photon at complementary channel is detected on the basis of $|\psi'\rangle$, the other photon at $n$-fold dephrasure channel automatically collapses to the state of $|\psi\rangle$. In such a way, any pure input state for the evolution can be remotely prepared~\cite{Yu2018}.

The preparation of the mixed input state is also feasible. For the mixed state $\rho$=$\lambda|H\rangle\langle H|^{\otimes n-1}+(1-\lambda)|V\rangle\langle V|^{\otimes n-1}$, we can rotate the state $|\Phi^+\rangle$ into  $\sqrt{\lambda}|H\rangle^{\otimes n}+\sqrt{1-\lambda}|V\rangle^{\otimes n})$ by a HWP, and then project the photon in the complementary channel into $\frac{1}{2}(|H\rangle\langle H|+|V\rangle\langle V|)$.

\begin{figure}[tb]
\includegraphics[width=0.47\textwidth]{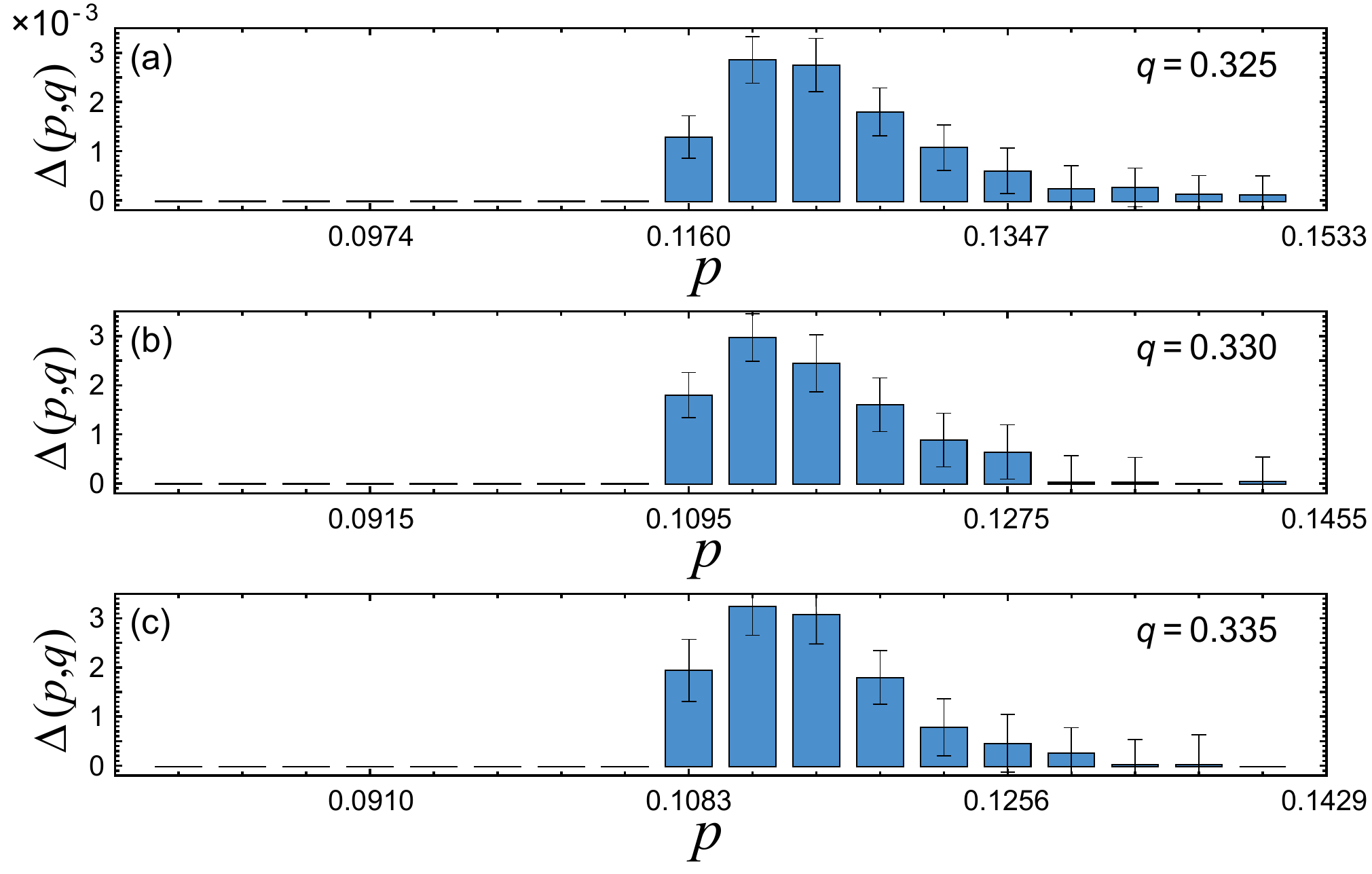}
\renewcommand{\thefigure}{S3}
\caption{\label{FigS3}
Detailed results of the superadditivity.}
\label{FigS3}
\end{figure}

\begin{figure}[tb]
\includegraphics[width=0.45\textwidth]{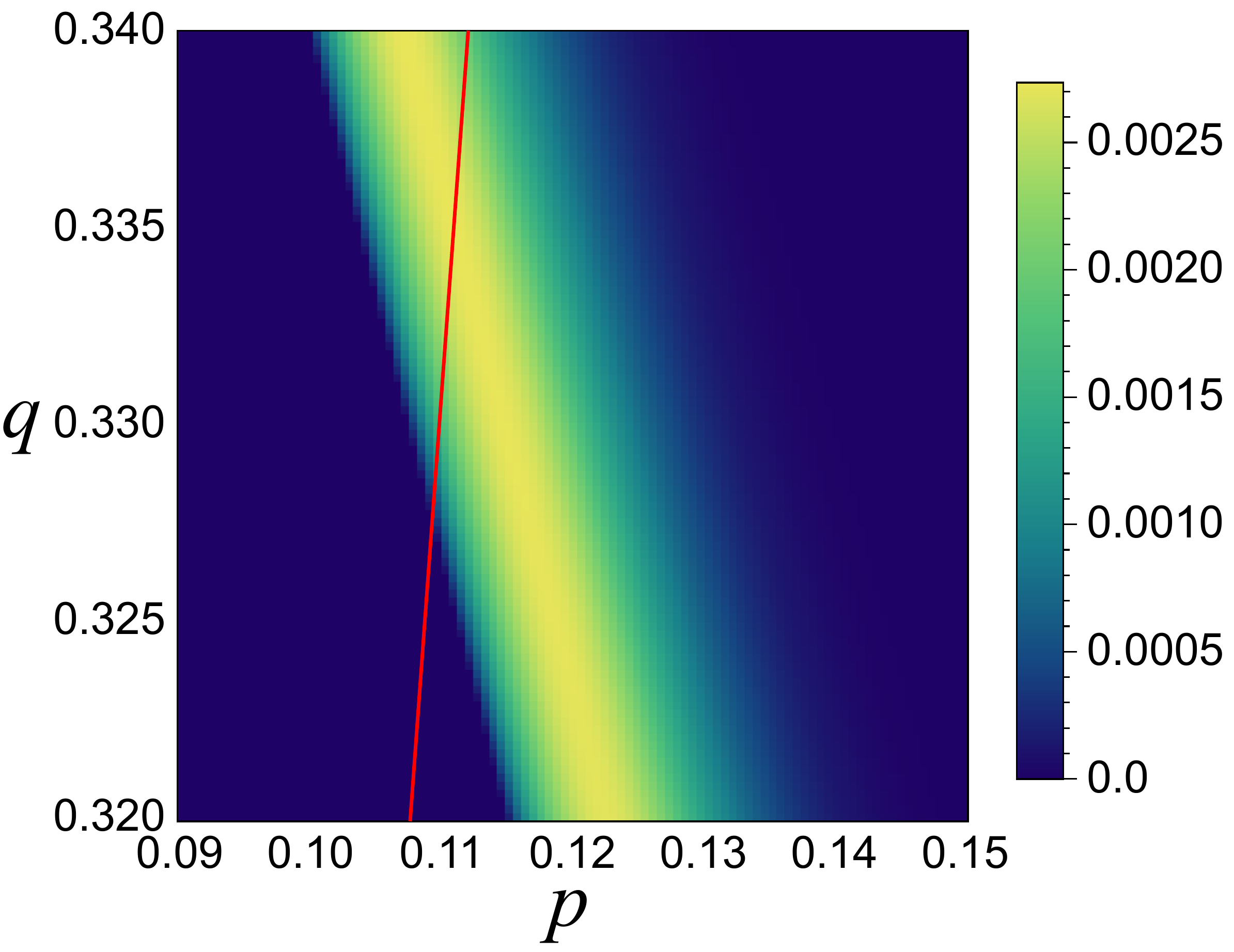}
\renewcommand{\thefigure}{S4}
\caption{\label{FigS4}
Numerical results of the $\Delta(p,q)$. The red line denote $q=3p$.}
\label{FigS4}
\end{figure}

\section{Determining the optimal proportion of noise parameters}
In order to clearly observe the superadditivite phenomenon, we need to find the optimal proportion of noise parameters $p$ and $q$. According to Ref.~\cite{Leditzky2018}, the superadditivity exhibits in the region $f_{1}(p)<q<f_{2}(p)$, where $f_{1}(p)=\frac{1-2p-2p(1-p)\ln\frac{1-p}{p}}{2-2p-2p(1-p)\ln\frac{1-p}{p}}$, $f_{2}(p)=\frac{(1-2p)^{2}}{1+(1-2p)^{2}}$. Fig. 3 in the main text shows some results about the superadditivity. In these results, we have determined the optimal $p$ and $q$ are located at $p\in[0.09,0.15]$, and $q\in[0.30,0.35]$. Thus, for determining the parameters more accurately, we need to do more measurements in this region.

In Fig. S3, we measure $\Delta(p,q)$ when $p=0.325$, $0.330$, and $0.335$. From the results, we find that the maximal $\Delta(p,q)$ appears at $p=0.11176$ when $q=0.335$. This evidence shows the optimal proportion of noise parameters is 2.997 (approximate to 3). In order to obtain a precise proportion, we also perform numerical simulations. Based on the experimental data at hand, we can control the noise parameters $p$ and $q$, in the range of $p\in[0.09,0.15]$, and $q\in[0.32,0.34]$. The corresponding result is shown in Fig. S4, and the optimal proportion is determined at $q=3p$.